\documentclass{svmult}

\usepackage{amsmath,amssymb,latexsym}

\usepackage{makeidx}         % allows index generation
\usepackage{graphicx}        % standard LaTeX graphics tool
\usepackage{multicol}        % used for the two-column index
\usepackage[bottom]{footmisc}% places footnotes at page bottom
\makeindex             % used for the subject index
\begin{document}

\title*{Modeling the non-Markovian, non-stationary scaling dynamics of 
financial markets}
\titlerunning{Non-Markovian, non-stationary scaling dynamics of a 
financial asset}
\author{
Fulvio Baldovin\inst{1}\,\inst{2}\and
Dario Bovina\inst{1}\,\inst{3}\and
Francesco Camana\inst{1}\,\inst{4}\and
Attilio L. Stella\inst{1}\,\inst{5}
}
\institute{
Dipartimento di Fisica,
Sezione INFN e CNISM, Universit\`a di Padova,
{\it Via Marzolo 8, I-35131 Padova, Italy}
\and
\texttt{baldovin@pd.infn.it}
\and
\texttt{bovina@pd.infn.it}
\and
\texttt{camana@pd.infn.it}
\and
\texttt{stella@pd.infn.it}
\footnote{Presented by A.L. Stella in a Talk at the
``Econophysics - Kolkata V'' International
Workshop, March 2010, Saha Institute of Nuclear Physics,
Kolkata, India.}
}

\maketitle

\begin{abstract}
A central problem of Quantitative Finance is that of formulating 
a probabilistic model of the time evolution of asset prices allowing 
reliable predictions on their future volatility. As in several 
natural phenomena, the predictions of such a model must be compared 
with the data of a single process realization in our records. In 
order to give statistical significance to such a comparison, 
assumptions of stationarity for some quantities extracted from the 
single historical time series, like the distribution of the returns 
over a given time interval, cannot be avoided. Such assumptions entail the 
risk of masking or misrepresenting non-stationarities of the underlying process,
and of giving an incorrect account of its correlations. Here we 
overcome this difficulty by showing that 
five years of daily Euro/US-Dollar trading records in the about three 
hours following the New York market opening, provide a rich 
enough ensemble of histories. The statistics  of this ensemble allows to propose and test 
an adequate model of the stochastic process driving the exchange 
rate. This turns out to be a non-Markovian, self-similar process with 
non-stationary returns. The empirical ensemble correlators are in
agreement with the predictions of this model, which is constructed 
on the basis of the time-inhomogeneous, anomalous scaling obeyed by 
the return distribution.
\end{abstract}

\date{\today}

\section{Introduction}
\label{sec_introduction}
The analysis of many natural and social phenomena is
hindered by the fact that one cannot replicate the dynamical 
evolution of the system under study. This may happen, for instance, 
for earthquakes \cite{scholz_1}, solar flares \cite{lu_1}, 
large eco-systems \cite{bjornstad_1}, and financial
markets \cite{cont_1}. If with a single time
series available we try to accommodate the historical data
within a stochastic process description, we must assume a priori 
the existence of some statistical quantities which remain stable over 
time \cite{cont_1}. This entitles us to sample their values at
different stages of the historical evolution, rather than at 
different instances of the process.
For example, in the analysis of historical series in Finance 
it is usual to assume the stationarity of the distribution of 
return fluctuations and hence to detect their statistical features 
through sliding time interval empirical sampling.
However, the plausible \cite{baldovin_1,stella_1,bassler_1,challet_1,andreoli_1} 
nonstationarity of these
fluctuations at intervals ranging from minutes to months would
drastically alter the relation between some of the 
stylized empirical facts detected in this way,
and the underlying stochastic process.
In order to identify the correct model, one has to overcome this 
difficulty. 
The breaking of time-translation invariance possibly signalled 
by increments non-stationarity would represent a challenge
in itself, being a genuine manifestation of dynamics out 
of equilibrium, like the aging properties observed in glassy 
systems \cite{aging_1}.

In order to detect the possible presence of nonstationarity 
at certain time-scales for the distribution of the increments, 
one would need to have access to many independent realizations of 
the same process, repeated under similar conditions.
Quite remarkably, high-frequency financial time-series offer an
opportunity of this kind, in which it is possible to directly sample
an ensemble of histories.
In Ref. \cite{bassler_1} it has been proposed that when considering
high-frequency EUR/USD exchange rate data as
recorded during the first three hours of the New York market 
activity, independent process realizations can tentatively be 
identified in the daily repetitions of the trading.
This gives the interesting possibility of estimating quantities related 
to \mbox{ensemble-,} rather than time-averages. Here we profit of this
opportunity by showing that a proper analysis of the statistical 
properties of this ensemble of histories naturally leads to
the identification and validation of an original stochatic
model of market evolution. The main idea at the basis of this model
is that the scaling properties of the return distribution are
sufficient to fully characterize the process in the time range
within which they hold.
The same type of model has been recently proposed by some of the
present authors to underlie
more generally the evolution of financial indices also in cases 
when only single realizations are available \cite{baldovin_1}. 
In those cases the
application of the model is less direct, and rests on suitable
assumptions about the relation between the stationarized empirical 
information obtainable from the historical series and the underlying
driving process.

An interesting feature of the model discussed here and in 
Refs. \cite{baldovin_1,stella_1},
is that the anomalous scaling of the return PDF enters in its
construction on the basis of a property of correlated stability   
which generalizes the stability of Gaussian PDF's under 
independent random variables summation. This correlated stability
was shown recently to allow the derivation of novel, constructive 
limit theorems for the PDF of sums of many strongly dependent random
variables obeying anomalous scaling \cite{limit}. 
In this perspective,
the model we present offers a valid alternative to more
standard models of Finance based on Gaussianity and independence. 
At the same time, the probabilistic framework provided by our 
modelization presents clear formal analogies and parallels with 
those standard models.

\section{An ensemble of histories based on the returns of
the EUR/USD exchange rate}
\label{sec_ensemble}
To address the above points, given the EUR/USD exchange
rate at time $t$ ($t$ measured in tens of minutes) 
after 9.00 am New York time, $S(t)$, let us
define the return in the interval $[t-T,t]$
as $R(t,T)\equiv\ln S(t)-\ln S(t-T)$, where $t=1,2,\ldots$, $t\geq T$.
By storing the daily repetitions of the returns from March 2000 to 
March 2005, we obtain an ensemble of $M=1,282$ realizations
$\left\{r^{l}(t,T)\right\}_{l=1,2,\ldots,M}$ of
the discrete-time stochastic process $R(t,T)$, with $t$ ranging in 
almost three hours after 9.00 am NY time, i.e., $1\leq t\leq17$.
Below, the superscript ``$e$'' labels quantities empirically
determined on the basis of this ensemble.
The first key observation is that the empirical second moment  
$m_2^e(t,1)\equiv \sum_{l=1}^M[r^l(t,1)]^2/M$ 
systematically decreases as a function of $t$
in the interval considered (see Fig. 1a).
This is a clear indication of return non-stationarity of the 
underlying process at this time scale.
In addition, an analysis of the nonlinear moments $m_\alpha^e$ of 
the total return $R(t,t)=\ln S(t)-\ln S(0)$ for $t\geq 1$,
\begin{equation}
m_\alpha^e(t,t)\equiv\frac{1}{M}\sum_{l=1}^M\left|r^l(t,t)\right|^\alpha,
\quad\alpha\in\mathbb{R}_+,
\end{equation} 
shows that such a nonstationarity is accompanied by an
anomalous scaling symmetry. Indeed, to a good approximation 
one finds $m_\alpha(t,t)\sim t^{\alpha D}$ in this range of $t$, 
where $D\simeq 0.364\ldots$ is essentially independent of $\alpha$ (Fig. 1b).
Accordingly, the ensemble histograms for the PDF's of aggregated returns in 
the intervals $[0,t]$, $p_{R(t,t)}$, are consistent with the scaling
collapse 
\begin{equation}
\label{eq_scaling}
t^D\;p_{R(t,t)}\left(t^D\;r\right)
=g(r)
\end{equation}
reported in Fig. 2. The scaling function $g$
identified by such collapse plot is manifestly non-Gaussian. It may 
also be assumed to be even to a good approximation\footnote{
We have detrended the data by subtracting from $r^l(t,T)$ 
the average value $\sum_{l=1}^M r^l(t,T)/M$. 
Data skewness can be shown
to introduce deviations much smaller than the statistical error-bars
in the analysis of the correlators. 
}.

\begin{figure}[t]
\label{fig_ns}
\vspace{0.5cm}
\centering
\begin{minipage}[t]{0.98\textwidth}
\includegraphics[width=0.49\textwidth]{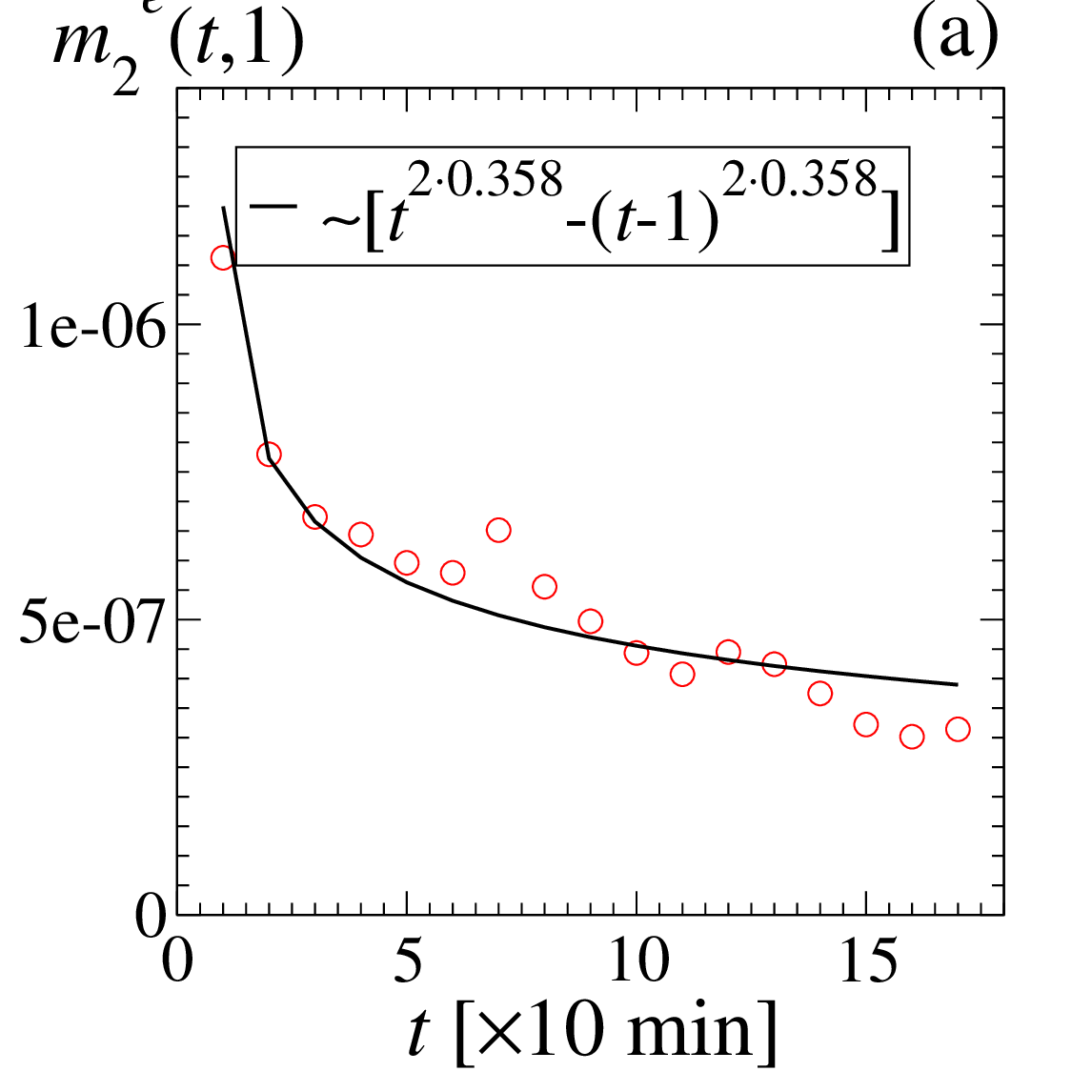}
\includegraphics[width=0.49\textwidth]{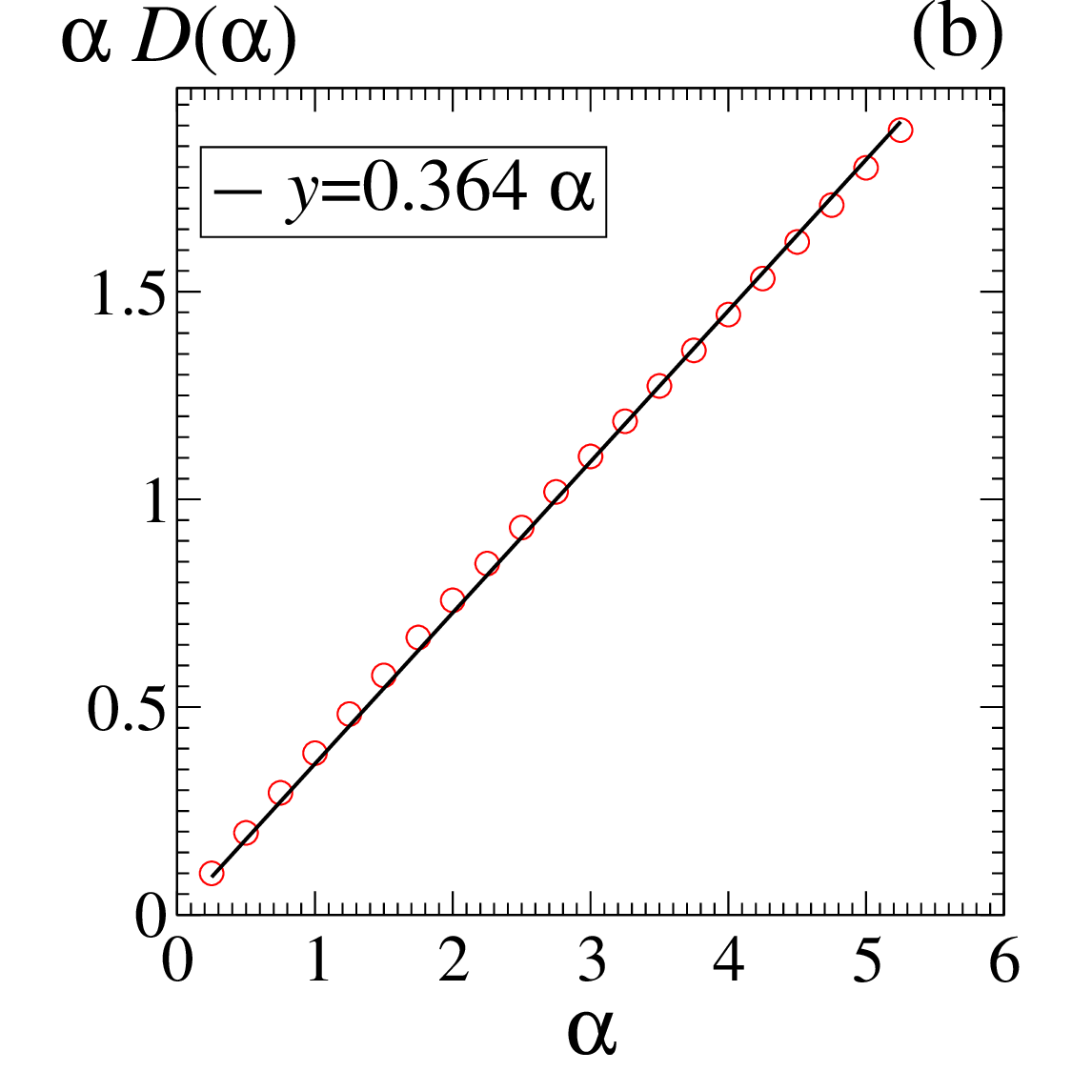}\\
$\;$\\
\vspace{0.5cm}
\includegraphics[width=0.98\textwidth]{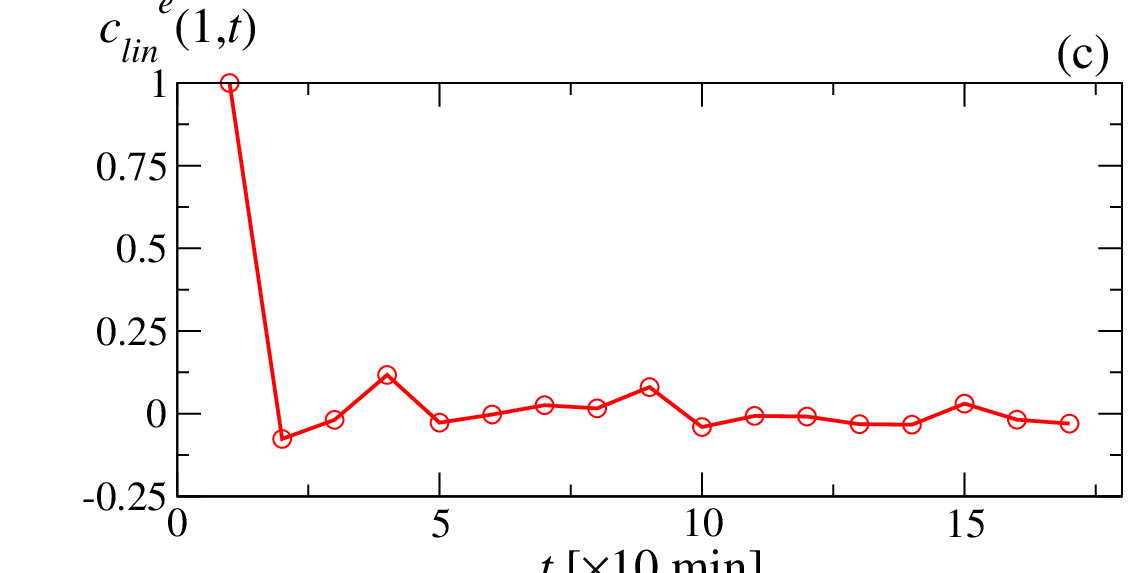}
\caption{
  Empirical ensemble analysis of the returns. 
  (a) The line is given by
  $\langle\sigma^2\rangle_\rho\;\left[t^{2D}-(t-1)^{2D}\right]$, 
  with $\langle\sigma^2\rangle_\rho=\langle r_1^2\rangle_p=2.3\cdot10 ^{-7}$ 
  and the best-fitted $D=0.358$.
  (b) Analysis according to the ansatz in Eq. (\ref{eq_scaling}). 
  The straight line characterizes a simple-scaling behavior
  with a best-fitted $D=0.364$. 
  (c) The linear correlation vanishes for non-overlapping returns. 
}
\end{minipage}
\end{figure}

To further simplify our formulas below, wherever appropriate we will
switch to the notations: $R_i \equiv R(i,1)$ and $r_i\equiv r(i,1)$.
Similarly $r^l_i \equiv r^l(i,1)$ will indicate the
$i$-th return on a $10\;\textrm{min}$-scale 
in the $l$-th history realization of our ensemble. 

An important empirical fact (Fig. 1c) is that the linear correlation between
returns
for non-overlapping intervals
\begin{equation}
c_{l i n}^e(1,n)\equiv
\frac{\frac{1}{M}\sum_{l=1}^M \left[r_1^l\;r_n^l\right]}
{\sqrt{m_2(1,1)\;m_2(n,1)}},
\end{equation}
with $n =2,\ldots$,
is negligible in comparison with
the correlation of the absolute values of the same returns. 
At this time scale also correlators of odd powers of 
a return with odd or even powers of another return are 
negligible. Only even powers of the returns are
strongly correlated.

\begin{figure}[t]
\label{fig_g}
\vspace{1.cm}
\centering
\begin{minipage}[t]{0.98\textwidth}
\includegraphics[width=0.98\columnwidth]{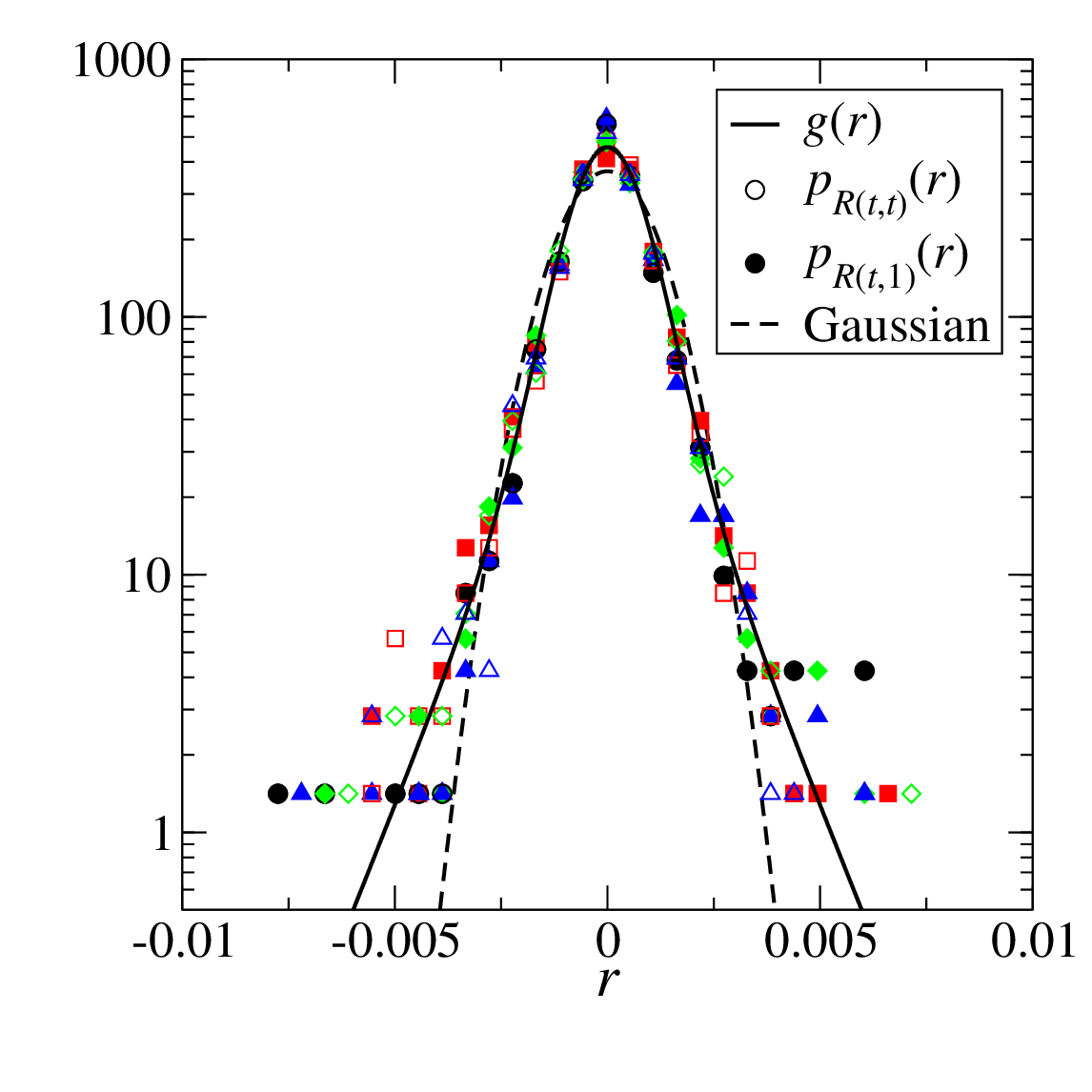}
\vspace{0.3cm}
\caption{
  Non-Gaussian scaling function $g$.
  Empty [full] symbols are obtained by rescaling $p_{R(t,t)}$
  [$p_{R(t,1)}$] according to Eq. (\ref{eq_scaling}) 
  [Eq. (\ref{eq_time_in})] for $t=1,5,10,17$.
}
\end{minipage}
\end{figure}

\section{Self-similar model process}
\label{sec_self_similar}
The empirical facts listed above already enable us to
suggest a very plausible model for the stochastic process
expected to generate the data. Both in physics and in Finance, a 
well established trend in modeling anomalous scaling is that of expressing 
the scaling functions,
like our $g$, as convex combinations of Gaussian PDF's with varying widths. 
This has clear mathematical advantages, since it is possible to express 
very general scaling functions with such convex combinations. In physics the 
representation in terms of mixtures of Gaussians often reflects the presence 
of some heterogeneity or polydispersity in the problem \cite{gheorghiu_1}. 
In Finance, 
the use of convex combinations of Gaussians to represent return
PDF's is naturally suggested by the fact that return time series
show a variety of more or less long intervals characterized by 
peculiar values of the volatility (volatility clustering). 
The idea that $p_{R(t,t)}$ can be represented as a mixture of Gaussians
of varying widths is suggested by the same basic motivations
which lead to the introduction of stochastic volatility 
models in Finance \cite{bouchaud_1,heston_1,gatheral_1,fouque_1}.
In the light of the empirical facts, such a representation of the 
scaling function in the PDF
of the aggregated return naturally suggests an adequate full 
modelization of the process generating the successive partial 
returns. Let us indicate by
$\rho(\sigma)$ a normalized, positive measure in $]0,+\infty[$
such that we can represent $g$ as:
\begin{equation}
\label{eq_mixture}
g(x)=\int_0^{+\infty} d\sigma \rho(\sigma) \frac{e^
  {-\frac{x^2}{2\sigma^2}}}{\sqrt{2 \pi \sigma^2}}.
\end{equation}
A suitable form of $\rho$ can be easily identified, e.g. by matching 
its moments with those of $g$, and by 
relating the large $\sigma$ behavior of $\rho(\sigma)$
with the large $|r|$ behavior of $g(r)$. 
For instance, $\rho$ may 
decay
as a power law at large $\sigma$'s if the moments of $g$ are expected
to be infinite above a given order. These conditions enable us to 
fix a number of parameters in $\rho$ such that the scaling function
in Eq. (\ref{eq_mixture}) fits the data in the empirical collapse 
in Fig. 2. As discussed below, in our case the set of data on 
which we can count to construct histograms of $g$ is relatively 
poor. So, our determinations of $\rho$ will be rather qualitative. 

Once identified $\rho$, more ambitiously we may try to use it 
for a weighted representation of the joint PDF's of
the successive elementary returns $R_i$, $i=1,2,\dots$ generated
in the process. Indeed, we may tentatively write the joint PDF 
of these returns in the following form:
\begin{equation}
\label{eq_joint_0}
p^{n}(r_1,r_2,\dots,r_{n})=
\int_0^{+\infty} d\sigma \rho(\sigma) \prod_{i=1}^{n}
\frac{
\exp\left({-\frac{r_i^2}{2\,a_i^2\,\sigma^2}}\right)
}{
\sqrt{2 \pi\,a_i^2 \,\sigma^2},
}
\end{equation}
with $n=1,2,\ldots,17$.
The coefficients $a_i$ in the last equation have to be chosen 
consistent with the non-stationarity of the elementary
returns reported in Fig. 1a and with the other statistical
properties of the elementary and aggregated returns discussed in
the previous section. It is straightforward
to realize that $\langle r_i^2\rangle_p =\langle\sigma^2\rangle_{\rho}\,a_i^2$,
while $\langle r_i\rangle_p=0$ and $\langle r_i r_j\rangle_p = 0$ for $i \neq j$,
where $\langle \cdot\rangle_p$ denotes averages with respect to the joint
PDF in Eq.(\ref{eq_joint_0}), 
whereas $\langle\cdot\rangle_\rho$ those with respect to the PDF
$\rho$.
Likewise, we immediately realize that odd-odd or odd-even correlators
of the $R_i$'s are strictly zero.
Assuming validity of Eq. (\ref{eq_joint_0}) 
means in first place that the $i$-dependence 
of $a_i$ must be chosen such to fit the values reported
in Fig. 1a. 
The choice of the $i$ dependence of $a_i$ 
must be also consistent with the simple scaling of the PDF 
of aggregated returns. Indeed, taking into account that
$R(t,t)=R_1 +R_2+ \dots+R_t$, for $t=1,2,\dots,17$,
Eq.(\ref{eq_joint_0}) implies that for the same $t$ values   
\begin{equation}
p_{R(t,t)}(r)=
\frac{g\left(r/\sqrt{a_1^2+a_2^2+\dots+a_{t}^2}\right)}{\sqrt{a_1^2+a_2^2+\dots+a_{t}^2}}. 
\end{equation}
Comparing this result with Eq. (\ref{eq_scaling}), we see that 
it is necessary to choose the $a_i$'s such that
$a_1^2+a_2^2+\dots+a_{t}^2 =t^{2D}$ in order to
be consistent with the empirical scaling in Eq. (\ref{eq_scaling}). 
This last requirement is
satisfied if we put 
\begin{equation}
\label{eq_ai}
a_i =\sqrt{i^{2D}-(i-1)^{2D}},\quad i=1,2,\ldots\;.
\end{equation}
A first problem is then to see whether this form of the $a_i$
coefficients is compatible with the $i$-dependence already implied
by the non-stationarity. 
\mbox{Eq. (\ref{eq_ai})} appears to be reasonably well compatible with 
the trend of the empirical mean square elementary returns 
$m_2(i,1)$. 
Indeed, given 
$\langle \sigma^2\rangle_{\rho}=\langle r_1^2\rangle_p=2.3\cdot10^{-7}$, 
the best fit in Fig. 1a
is obtained with $D=0.358\ldots$ in
the expression for $\langle r_i^2\rangle_p$. The expectation value of $\sigma^2$ is
with respect to the $\rho$ entering the integral representation (3) 
already chosen for $g$. Remarkably, the value of $D$ is very close 
to the estimate of $D$ obtained above through the analysis of the
moments of $p_{R(t,t)}$.

Summarizing, Eq.(\ref{eq_joint_0}) and the above conditions on
the $a_i$'s define a non-Markovian stochastic process with linearly 
uncorrelated increments and a PDF of returns satifying 
a time inhomogeneous scaling of the form:
\begin{equation}
p_{R(t,T)}(r)=
\frac{1}{\sqrt{t^{2D}-(t-T)^{2D}}}\;\;
g\left(\frac{r}{\sqrt{t^{2D}-(t-T)^{2D}}}\right),
\label{eq_time_in}
\end{equation}
where both $t$ and $T$ are understood to be integer multiples 
of the $\rm{10}$ minutes unit. 
In Fig. 2 it is shown that the data collapse of
both $p_{R(t,t)}$ and $p_{R(t,1)}$ are indeed compatible with 
the same non-gaussian PDF $g$.

From the point of view of probability theory, the structure
of our process in Eq.(\ref{eq_joint_0}) rests on a stability property
for PDF's of sums of dependent random variables \cite{limit}. Indeed, if we
indicate by $\tilde{p}^{n}(k_1,k_2,...k_n)$ the Fourier transform 
(characteristic function) of the joint
PDF of the first $n$ returns ($1\leq n \leq 17$), a direct 
calculation yields
\begin{equation}
\tilde{p}^{n}(k,k,...,k)=\tilde{p}^{1}\left(n^DK\right)
\end{equation}
and
\begin{equation}
\tilde{p}^{n}(0,..,k_i,..,0)=\tilde{p}^{1}(a_i k_i),\quad i=1,\dots,n.
\end{equation}
For $D=1/2$ these relations have the the same form as those holding 
in the case of independent variables, when $\tilde{p}^n(k_1,\dots,k_n)=
\tilde{p}^1(k_1)\,\tilde{p}^1(k_2)\,\dots\,\tilde{p}^1(k_n)$,
and $\tilde{p}^1$ is a Gaussian characteristic function.
However, even for $D=1/2$ 
a general $\rho(\sigma)$ implies dependence of the
$R_i$'s.
To recover the independent case one needs further to choose 
$\rho(\sigma)=\delta(\sigma-\sigma_0)$. Thus, the superposition 
of independent Gaussian processes with different $\sigma$'s
in Eq.(\ref{eq_joint_0}) implies an extension of the basic stability
properties of the independent Gaussian variables case to 
the dependent case. This extension also allows to
derive limit theorems for the anomalous scaling 
of sums of many dependent random variables \cite{limit}.

\section{Correlations structure}
\label{sec_corr}
As discussed above, the identification of $\rho$ may be used to
reconstruct the joint PDF of the returns $R_i$'s as in
Eq. (\ref{eq_joint_0}).  
In this section we elaborate further on this point, by performing a
detailed comparison between model predictions (based on an explicit
expression for $\rho$) and empirical determinations
of various two-point correlators.

Considering the data collapse of both
$p_{R(t,t)}$ and $p_{R(t,1)}$ in Fig. 2, we propose the
following functional form for $\rho$ (see also \cite{limit}):
\begin{equation}
\rho(\sigma)=A\;
\frac{\sigma^{\gamma}}{d+\sigma^\delta},
\quad \sigma\in[\sigma_{min},+\infty[,
\quad 0<\gamma<\delta,
\end{equation}
where $A$ is a normalization factor, and $d>0$ is a parameter
influencing the width of the distribution $g$. 
Notice that $\rho(\sigma)\sim\sigma^{-(\delta-\gamma)}$ for $\sigma\gg1$.
The rational behind this choice for $\rho$
is that one can use the exponents $\gamma,\delta$ to reproduce the large $|x|$ 
behavior of $g(x)$, and then play with the other parameters to obtain a
suitable fit of the scaling function, for instance the one
reported in Fig. 2.

The first two-point correlator we consider in our analysis is 
\begin{equation}
\kappa_{\alpha,\beta}(1,n)\equiv
\frac{
\langle\left |R(1,1)\right |^\alpha\;
\left |R(n,1)\right |^\beta\rangle
}
{
\langle\left |R(1,1)\right |^\alpha\rangle_p\;
\langle\left |R(n,1)\right |^\beta\rangle_p
}
=
\frac{
\langle\left |r_1\right |^\alpha\;
\left |r_n\right |^\beta\rangle_p
}
{
\langle\left |r_1\right |^\alpha\rangle_p\;
\langle\left |r_n\right |^\beta\rangle_p
}, 
\end{equation}
with $n>1$, and $\alpha,\beta\in\mathbb R_+$.
A value $\kappa_{\alpha,\beta}\neq1$ means that returns on non-overlapping
intervals are dependent. 
Using Eq. (\ref{eq_joint_0}) it is possible to
express a general many-return correlator in terms of the moments of
$\rho$.
For example, from Eq. (\ref{eq_joint_0}) we have 
\begin{equation}
\langle\left |r_1\right |^\alpha\;
\left |r_n\right |^\beta\rangle_p=
B_\alpha\,B_\beta\;\;a_1^\alpha\,a_n^\beta\;
\langle\sigma^{\alpha+\beta}\rangle_\rho,
\label{eq_corr_1}
\end{equation}
with 
\begin{equation}
B_\alpha\equiv\int_{-\infty}^{+\infty}d r\;|r|^\alpha\;
\frac{e^{-r^2/2}}{\sqrt{2\pi}}.
\end{equation}
We thus obtain 
\begin{equation}
\kappa_{\alpha,\beta}(1,n)=
\frac{\langle\sigma^{\alpha+\beta}\rangle_\rho}
{\langle\sigma^\alpha\rangle_\rho\;
\langle\sigma^\beta\rangle_\rho}=
\frac{B_\alpha B_\beta}{B_{\alpha+\beta}}
\frac{\langle |r_1|^{\alpha+\beta}\rangle_p}
{\langle |r_1|^\alpha\rangle_p\;
\langle |r_1|^\beta\rangle_p}.
\label{eq_corr_3}
\end{equation}
Two model-predictions in Eq. (\ref{eq_corr_3}) are: (i) Despite the
non-stationarity of the increments $R_i$'s,
$\kappa_{\alpha,\beta}(1,n)$ is independent of $n$;
(ii) The correlators are symmetric, i.e., $\kappa_{\alpha,\beta}-\kappa_{\beta,\alpha}=0$.

\begin{figure}[t]
\label{fig_corr_increments}
\vspace{1.cm}
\centering
\begin{minipage}[t]{0.98\textwidth}
\includegraphics[width=0.98\columnwidth]{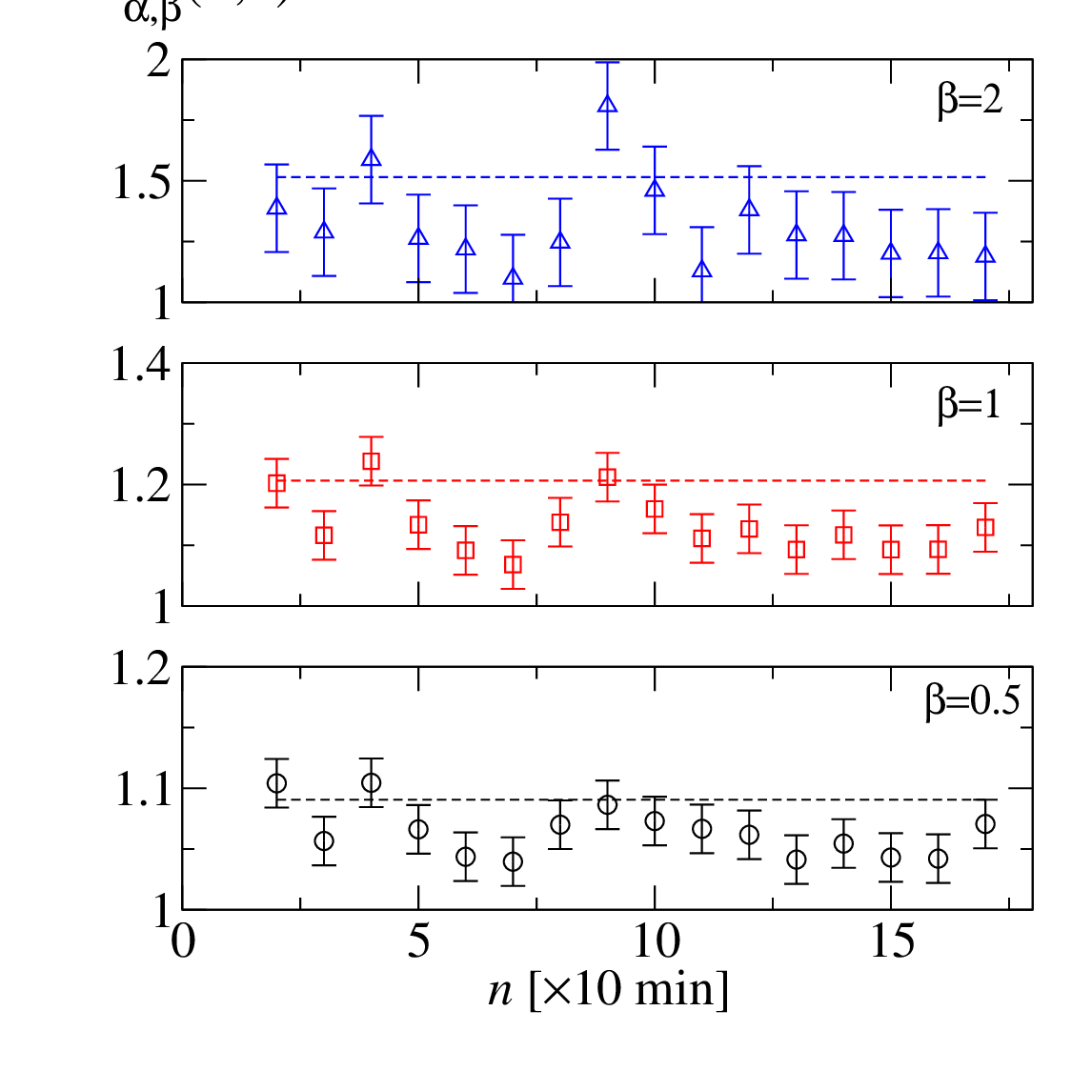}\\
\caption{
  Constancy of $\kappa^e_{\alpha,\beta}$. Dashed lines are
  model-predictions. 
}
\end{minipage}
\end{figure}

We can now compare the theoretical
prediction of the model 
for $\kappa_{\alpha,\beta}(1,n)$, Eq. (\ref{eq_corr_3}), with the empirical counterpart
\begin{equation}
\kappa^e_{\alpha,\beta}(1,n)\equiv
\frac{
\sum_{l=1}^M \left[\left |r_1^l\right |^\alpha\;
\left |r_n^l\right |^\beta\right]
}
{
\frac{1}{M}\sum_{l=1}^M \left |r_1^l\right |^\alpha\;
\sum_{l=1}^M \left |r_n^l\right |^\beta
}, 
\end{equation}
which we can calculate from the EUR/USD dataset.
Notice that once $\rho$ is fixed to fit the one-time statistics
in Fig. 2, in this comparison we do not have any
free parameter to adjust. 
Also, since our ensemble is restricted to $M=1,282$ realizations only,
large fluctuations, especially in two-time statistics, are to be
expected. 

Fig. 3 shows that indeed non-overlapping
returns are strongly correlated in the about three hours following
the opening of the trading session, since $\kappa^e_{\alpha,\beta}\neq1$.
In addition, 
the constancy of
$\kappa^e_{\alpha,\beta}$ is clearly suggested by the empirical data.   
In view of this constancy, we can assume as error-bars for 
$\kappa^e_{\alpha,\beta}$ the standard deviations of the sets 
$\left\{\kappa^e_{\alpha,\beta}(1,n)\right\}_{n=2,3,\ldots,17}$.
The empirical values for $\kappa^e_{\alpha,\beta}$ are also in agreement
with the theoretical predictions for $\kappa_{\alpha,\beta}$ based on our
choice for $\rho$. In this and in the following comparisons it should
be kept in mind that, although not explicitly reported in the plots, 
the uncertainty in the identification of $\rho$
of course introduces an uncertainty in the
model-predictions for the correlators.  

\begin{figure}[t]
\label{fig_alpha_beta}
\vspace{1.cm}
\centering
\begin{minipage}[t]{0.98\textwidth}
\includegraphics[width=0.98\columnwidth]{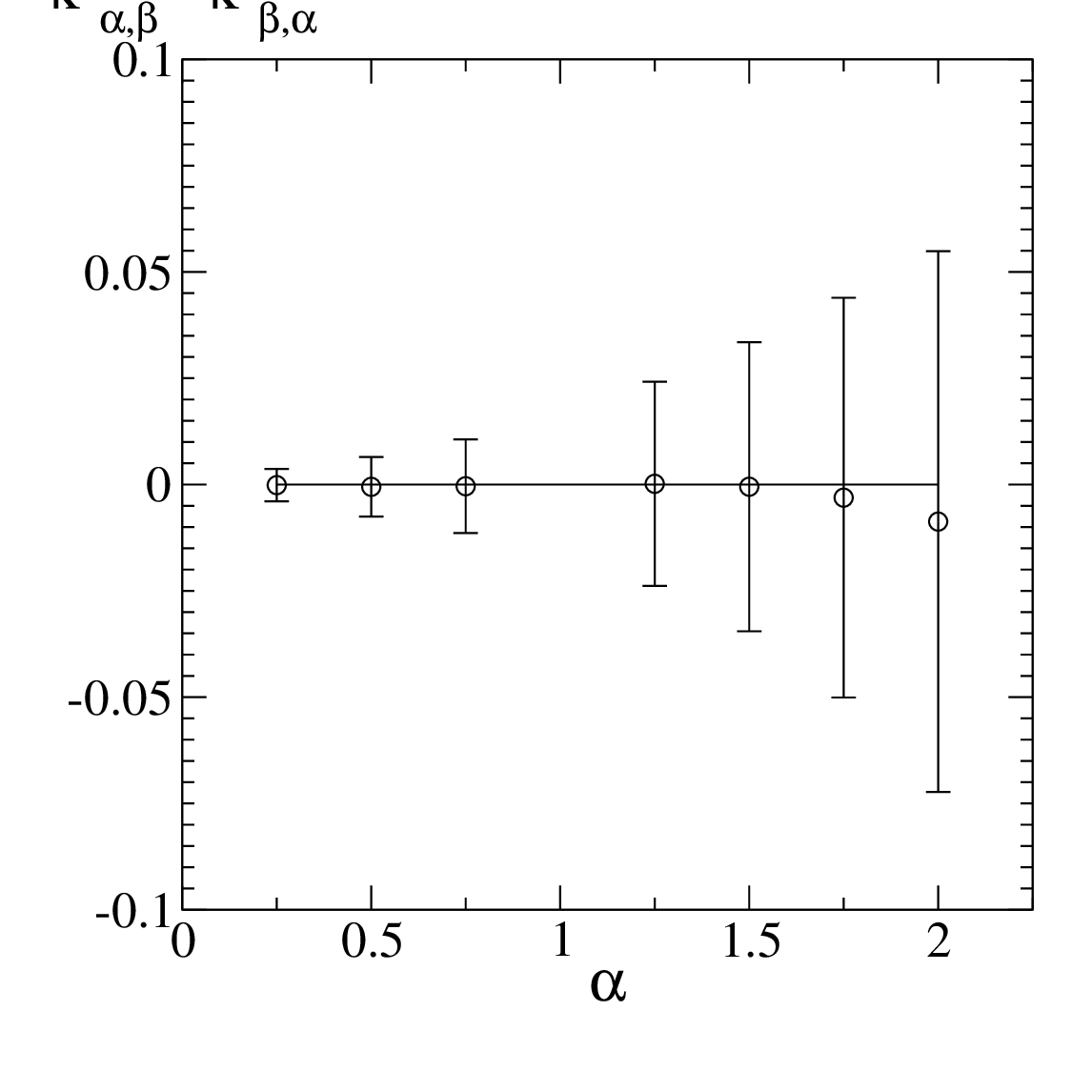}
\vspace{0.3cm}
\caption{
  Symmetry of $\kappa^e_{\alpha,\beta}$. Error-bars are determined as in
  Fig. 3. 
}
\end{minipage}
\end{figure}

In Fig. 4 
we report that also the symmetry 
$\kappa_{\alpha,\beta}=\kappa_{\beta,\alpha}$ is emiprically verified for the
EUR/USD dataset.
The validity of this symmetry for a process with non-stationary
increments like the present one is quite remarkable. 

\begin{figure}[t]
\label{fig_autocorrelation}
\vspace{1.cm}
\centering
\begin{minipage}[t]{0.98\textwidth}
\includegraphics[width=0.98\columnwidth]{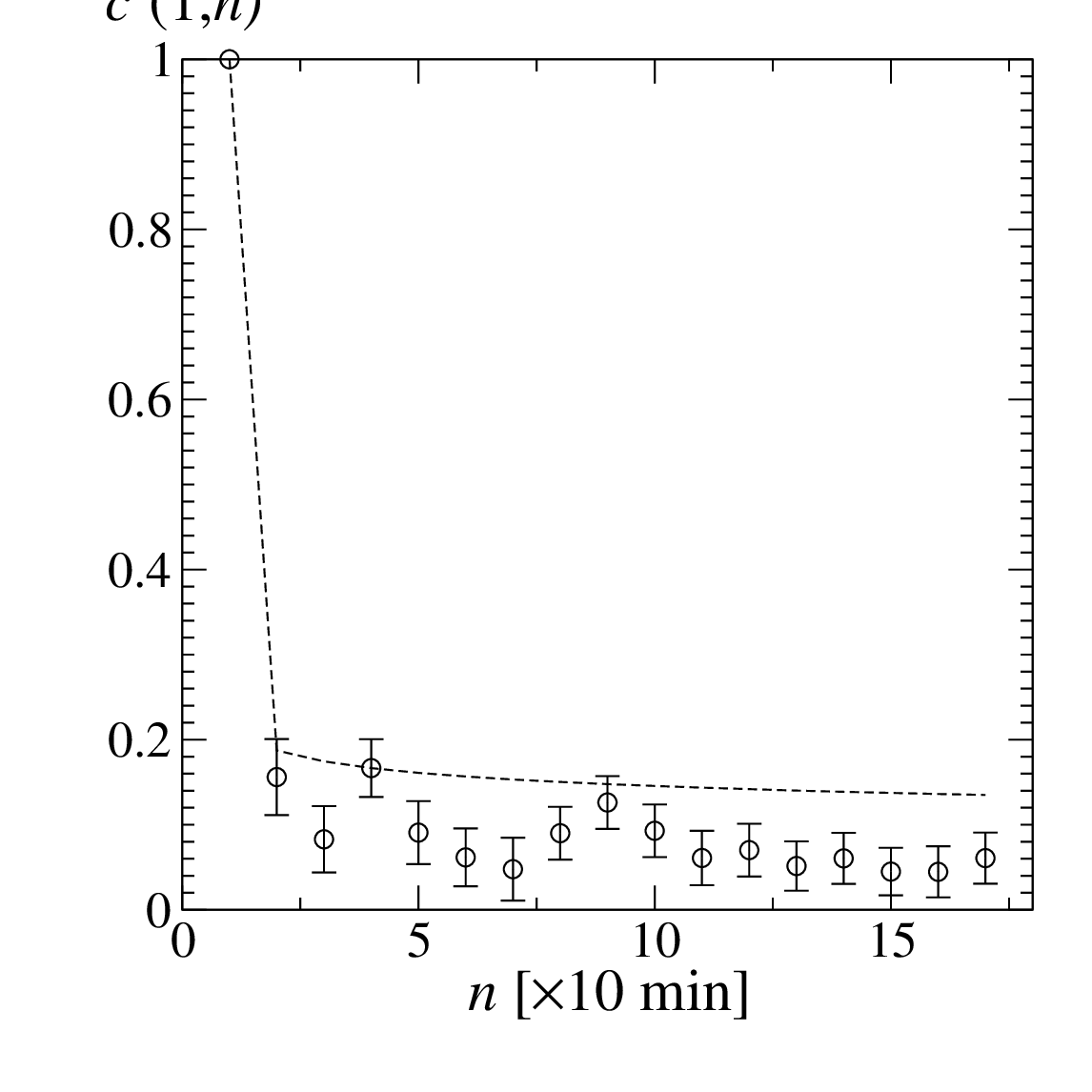}
\caption{
  Volatility autocorrelation. Dashed line is the 
  model-prediction. 
}
\end{minipage}
\end{figure}

\begin{figure}
\centering
\vspace{1.cm}
\begin{minipage}[t]{0.98\textwidth}
\label{fig_process}
\includegraphics[width=0.98\columnwidth]{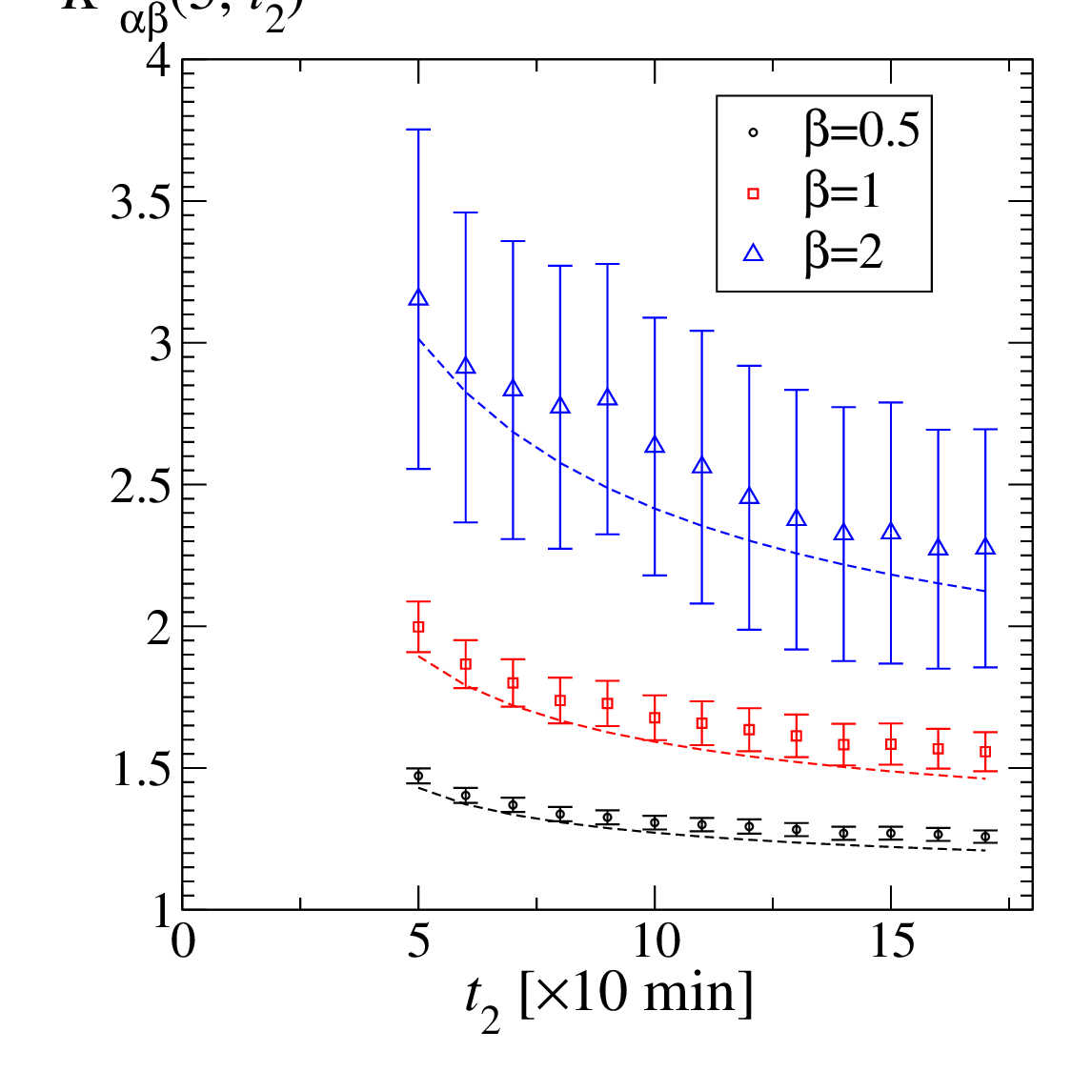}
\caption{
  Correlators $K^e_{\alpha,\beta}$. Dashed lines are
  model-predictions. 
}
\end{minipage}
\end{figure}

A classical indicator of strong correlations in financial data is the
volatility autocorrelation, defined as
\begin{equation}
c(1,n)\equiv
\frac{
\langle\left |r_1\right |\;
\left |r_n\right |\rangle_p
-
\langle\left |r_1\right |\rangle_p\;
\langle\left |r_n\right |\rangle_p
}
{
\langle\left |r_1\right |^2\rangle_p\;
-
\langle\left |r_1\right |\rangle_p^2
}. 
\end{equation}
In terms of the moments of $\rho$, 
through Eq. (\ref{eq_corr_1}) we have the following expression for 
$c$ :
\begin{equation}
c(1,n)=
\frac{
B_1^2\;a_1\;a_n\;
\left[
\langle\sigma^2\rangle_\rho
-
\langle\sigma\rangle_\rho^2
\right]
}
{
a_1^2\;
\left[
B_2\;
\langle\sigma^2\rangle_\rho
-
B_1^2
\langle\sigma\rangle_\rho^2
\right]
}. 
\end{equation}
Unlike $\kappa_{\alpha,\beta}$, $c$ 
is not constant in $n$. 
The comparison with the empirical volatility autocorrelation,
\begin{equation}
c^{e}(1,n)\equiv
\frac{
\sum_{l=1}^M \left[\left |r_1^l\right |\;
\left |r_n^l\right |\right]
-
\frac{1}{M}
\sum_{l=1}^M \left |r_1^l\right |\;
\sum_{l^\prime=1}^M \left |r_n^{l^\prime}\right |\;
}
{
\sum_{l=1}^M \left |r_1^l\right |^2
-
\frac{1}{M}
\sum_{l=1}^M \left |r_1^l\right |\;
\sum_{l^\prime=1}^M \left |r_1^{l^\prime}\right |\;
},
\end{equation}
yields a substantial agreement (See Fig. 5). 
The error-bars in Fig. 5 are obtained by
dynamically generating many ensembles of $M=1,282$ realizations each, according to
Eq. (\ref{eq_joint_0}) with our choice for $\rho$, and taking the
standard deviations of the results. 
Again, the uncertainty associated to the theoretical prediction for
$c$ is not reported in the plots
Problems concerning the numerical simulation of processes like the
one in Eq. (\ref{eq_joint_0}) are discussed in Ref. \cite{limit}. 

A further test of our model 
can be made by analyzing, 
in place of those of the increments,
the non-linear correlators of $R(t,t)$, with
varying $t$. 
To this purpose, let us define
\begin{equation}
K_{\alpha,\beta}(t_1,t_2)\equiv
\frac{
\langle\left |R(t_1,t_1)\right |^\alpha\;
\left |R(t_2,t_2)\right |^\beta\rangle
}
{
\langle\left |R(t_1,t_1)\right |^\alpha\rangle\;
\langle\left |R(t_2,t_2)\right |^\beta\rangle
},
\end{equation}
with $t_2\geq t_1$.
Model calculations similar to the previous ones give, from
Eq. (\ref{eq_joint_0}),
\begin{equation}
K_{\alpha,\beta}(t_1,t_2)=
\frac{B_{\alpha,\beta}^{(2)}(t_1,t_2)}{t_1^{\alpha D}t_2^{\beta D}B_{\alpha+\beta}}
\frac{\langle\left |r_1\right |^{\alpha+\beta}\rangle_p}
{\langle\left |r_1\right |^\alpha\rangle_p\;
\langle\left |r_1\right|^\beta\rangle_p},
\label{eq_corr_process}
\end{equation}
where
\begin{eqnarray}
&B_{\alpha,\beta}^{(2)}(t_1,t_2)\equiv
\int_{-\infty}^{+\infty}d r_1\;|r_1|^\alpha\;
\frac{\exp\left(-r_1^2/\left(2t_1^{2D}\right)\right)}
{\sqrt{2\pi t_1^{2D}}}&\nonumber\\
&\int_{-\infty}^{+\infty}d r_2\;|r_2|^\beta\;
\frac{\exp\left[-(r_1-r_2)^2/\left(2t_2^{2D}-2t_1^{2D}\right)\right]}
{\sqrt{2\pi \left(t_2^{2D}-t_1^{2D}\right)}}.&
\end{eqnarray}
According to Eq. (\ref{eq_corr_process}), $K_{\alpha,\beta}$ is
now identified by both $\rho$ and $D$. 
Moreover, it explicitly depends on $t_1$ and $t_2$.
The comparison between Eq. (\ref{eq_corr_process}) and
the empirical quantity 
\begin{equation}
K^e_{\alpha,\beta}(t_1,t_2)\equiv
\frac{
\sum_{l=1}^M \left[\left |r^l(t_1,t_1)\right |^\alpha\;
\left |r^l(t_2,t_2)\right |^\beta\right]
}
{
\frac{1}{M}\sum_{l=1}^M \left[\left |r^l(t_1,t_1)\right |^\alpha\right]\;
\sum_{l=1}^M \left[\left |r^l(t_2,t_2)\right |^\beta\right]
}, 
\end{equation}
reported in Fig. 6 
(the error-bars are determined as in Fig. 5)
supplies thus an additional 
validation of our model.

\section{Conclusions}
In the present work we addressed the problem of
describing the time evolution of financial assets in a case 
in which one can try to compare the predictions of
the proposed model with a relatively rich ensemble of
history realizations. Besides the fact that considering 
the histories at disposal for the EUR/USD exchange rate
as a proper ensemble amounts to a main working assumption, a clear 
limitation of such an approach is the relative poorness of
the ensemble itself. Indeed, the simulations of our model
suggest that in order to reduce substantially the 
statistical fluctuations one should dispose of ensembles larger 
by at least one order of magnitude.

In spite of these limitations, we believe that the non-Markovian model
we propose \cite{baldovin_1,stella_1,limit} 
is validated to a reasonable extent by the analysis
of the data, especially those pertaining to the various
correlators we considered. In this respect it is important to
recall that the first proposal of the time inhomogeneous
evolution model discussed here has been made in a study of
a single, long time series of the DJI index in Ref. \cite{baldovin_1}.
In that context, the returns time inhomogeneity, Eq. (\ref{eq_time_in}),
was supposed to underlie the stationarized information
provided by the empirical PDF of the returns. This assumption
allowed there to give a justification of several stylized 
facts, like the scaling and multiscaling of the empirical 
return PDF and the power law behavior in time of the return 
autocorrelation function. We believe that the results
obtained in the present report, even if pertaining to a different
time-scale (tens of minutes in place of days),
constitute an interesting further argument in favor of a 
general validity of the model.

The peculiar feature of this model is that of focussing
on scaling and correlations as basic, closely connected
properties of assets evolution. This was strongly inspired by
what has been learnt in the physics of complex systems in the
last decades \cite{kadanoff_1,kadanoff_2,lasinio_1}, 
where methods like the renormalization
group allowed for the first time systematic treatments of
these properties \cite{stella_1}. 
At the same time, through
the original probabilistic parallel mentioned in 
\mbox{Section \ref{sec_self_similar}}, 
our model maintains an interesting direct contact with the 
mathematics of standard formulations based on Brownian motion, 
of wide use in Finance. This last feature is very interesting 
in the perspective of applying our model to problems of 
derivative pricing \cite{bouchaud_1,heston_1,gatheral_1,fouque_1,black_1}.

{\bf Acknowledgments}\\
We thank M. Caporin for useful discussions. 
This work is supported by 
``Fondazione Cassa di Risparmio di Padova e Rovigo'' within the 
2008-2009 ``Progetti di Eccellenza'' program.

\printindex
\end{document}